\documentclass[useAMS,usenatbib]{mn2e}
\usepackage{graphicx}
\usepackage{epsf}
\usepackage{amsmath}
\newcommand{\apj}{ApJ}
\newcommand{\apjl}{ApJL}

\newcommand{\araa}{ARA\&A}
\newcommand{\mnras}{MNRAS}

\newcommand{\erg}{{\rm erg}}
\newcommand{\Hz}{{\rm Hz}}

\newcommand{\cm}{{\rm cm}}

\newcommand{\hmpc}{h^{-1}{\rm Mpc}}

\newcommand{\kms}{\;{\rm km}\,{\rm s}^{-1}}
\newcommand{\msun}{M_{\odot}}

\newcommand{\MUV}{M_{\rm UV}}

\newcommand{\Mmin}{M_\mathrm{min}}

\begin{document}

\title[The Minimum Halo Mass for Star Formation]{The Minimum Halo Mass for Star Formation at $z=$6--8} 

\author[K.\ Finlator et al.]{
\parbox[t]{\textwidth}{\vspace{-1cm}
Kristian Finlator$^{1,8,11}$, 
Moire K.~M.~Prescott$^{1,8}$,
B.\ D.\ Oppenheimer$^{4}$,
Romeel Dav\'e$^{5,6,7}$,
E.\ Zackrisson$^3$,
R.\ C.\ Livermore$^{10}$,
S.\ L.\ Finkelstein$^{10}$,
Robert Thompson$^{9}$,
Shuiyao Huang$^{2}$
}
\\\\$^1$ New Mexico State University, Las Cruces, NM, USA
\\$^2$ University of Massachussetts, Amherst, MA, USA
\\$^3$ Department of Physics and Astronomy, Uppsala University, 751 20 Uppsala, Sweden
\\$^4$ CASA, Department of Astrophysical and Planetary Sciences, University of Colorado, 389-UCB, Boulder, CO 80309, USA
\\$^5$ University of the Western Cape, Bellville, Cape Town 7535, South Africa
\\$^6$ South African Astronomical Observatories, Observatory, Cape Town 7525, South Africa
\\$^7$ African Institute for Mathematical Sciences, Muizenberg, Cape Town 7545, South Africa
\\$^8$ Dark Cosmology Centre, Niels Bohr Institute, University of Copenhagen, Copenhagen, Denmark
\\$^9$ NCSA, University of Illinois, Urbana-Champaign, IL 61820
\\$^{10}$ The University of Texas at Austin, 2515 Speedway, Stop C1400, Austin, TX 78712, USA
\\$^{11}$ finlator@nmsu.edu
\author[The Minimum Mass for Star Formation at $z=$6--8]{
K.\ Finlator,
B.\ D.\ Oppenheimer,
R.\ Dav\'e,
E.\ Zackrisson,
R.\ Thompson,
\& S.\ Huang,
}
}

\maketitle

\begin{abstract}
Recent analysis of strongly-lensed sources in the Hubble Frontier Fields 
indicates that the rest-frame UV luminosity function of galaxies at $z=$6--8 
rises as a power law down to $\MUV=-15$, and possibly as faint as -12.5.  
We use predictions from a cosmological 
radiation hydrodynamic simulation to map these luminosities onto physical space,
constraining the minimum dark matter halo mass and stellar mass that the Frontier 
Fields probe.  While previously-published theoretical studies have suggested or assumed that 
early star formation was suppressed in halos less massive than $10^9$--$10^{11}\msun$, 
we find that recent observations demand vigorous star formation in halos at 
least as massive as
(3.1, 5.6, 10.5)$\times10^9\msun$ at $z=(6,7,8)$.  Likewise, we find that 
Frontier Fields observations probe down to stellar masses of 
(8.1, 18, 32)$\times10^6\msun$; that is, they are observing the 
likely progenitors of analogues to Local Group dwarfs such as Pegasus and M32.  
Our simulations yield somewhat different constraints than two complementary 
models that have been invoked in similar analyses, emphasizing the need for 
further observational constraints on the galaxy-halo connection.
\end{abstract}


\section{Introduction} \label{sec:intro}
A key question regarding the way in which dark matter halos grow galaxies 
is the minimum mass of a dark matter halo $\Mmin$ that can both retain its 
gas and condense it efficiently onto a galaxy.  Locally, this question is central to both 
the ``missing satellites" and the ``too big to fail" problems~\citep{kly99,boy11}.
In the context of the reionization epoch ($z\geq6$), it arises because
of the possible role of faint galaxies in driving the growth of the 
nascent ultraviolet ionizing background (UVB): the steep observed slope 
of the UV luminosity function's (LF) 
faint end gives rise to a luminosity density that diverges if integrated to
arbitrarily faint luminosities.  Hence faint galaxies could well have driven 
reionization and dominated the UVB, but only if the limiting luminosity out 
to which the LF continues as a power law is quite faint~\citep{rob13,fink12,kuh12b}.  
As a minimum luminosity would imply a minimum halo mass, measurements of the
former can be invoked as a constraint on the latter.

This, in turn, would constrain a number of physical processes that can
regulate gas cooling, star formation, and feedback in low-mass halos.  
For example, ``Jeans Filtering" prevents gas from being accreted by 
dark matter halos whose virial temperature is less than the (appropriately 
time-averaged) temperature of the ambient intergalactic medium 
(IGM;~\citealt{efs92,qui96,gne00,oka08}).  While the effect is expected
to play a role even in the absence of reionization~\citep{nao09}, 
idealized simulations 
predict that the latent heat from photoionization completely halts gas 
accretion in halos with circular velocities $V_\mathrm{circ}$
below $30\kms$ but is negligible for $V_\mathrm{circ}>75\kms$~\citep{tho96}. 
Similarly, three-dimensional simulations indicate that it suppresses the gas 
reservoirs of halos less massive than $3\times10^8\msun$ 
($V_\mathrm{circ}<26\kms$) at $z=6$~\citep{oka08,fin12}.
This idea has motivated extended reionization models in which star
formation is assumed not to occur in reionized regions in halos less massive 
than $1\mbox{--}2\times10^9\msun$~\citep{ili07,alv12,mes08}.  
Building further on this idea,~\citet{bou10} proposed that gas in halos 
less massive than $10^{11}\msun$ does not accrete efficiently onto the 
central galaxy owing to photoionization feedback from hot stars at $z>6$.
They showed that this assumption naturally allows a simple model to 
reproduce measurements of galaxy downsizing at $z\leq2$.  Note that 
these theoretical studies invoke similar physics but assume threshold 
masses that vary by two orders of magnitude, clearly motivating 
the need for observational constraints.

An additional effect that is related to Jeans suppression is photoevaporation:
During the reionization epoch, ionization fronts likely evaporated gas that was 
bound to minihalos ($< 10^{7\mbox{--}8}\msun$;\citealt{sha04}), suppressing 
further star formation below
this mass range.  Even in the absence of a UVB, halos whose virial temperature 
is less than $10^4$K cannot cool and condense their gas through collisional 
excitation of neutral hydrogen; they are dependent on molecular hydrogen
formation cooling, which is relatively inefficient.  These effects have been 
modeled in high-resolution, ab-initio numerical simulations, leading to 
predictions that the reionization-epoch LF flattens for luminosities 
$\MUV < -12$~\citep{wis14}.  

Another source of suppression is galactic outflows, which are closely
associated with vigorous star formation~\citep{vei05}.  A variety of
theoretical models indicate that outflows are more efficient at removing 
gas from galaxies and suppressing their growth when they live in low-mass 
halos~\citep{dek86,hec02,mur05,mura15,chr16}.  They may even introduce a
characteristic scale below which suppression is particularly 
efficient~\citep{dek86,mura15}.  

Finally, recent models in which the local star formation rate density is
computed from the local density of molecular hydrogen (as opposed to the 
total gas density) predict that galaxy growth is suppressed in dark matter 
halos less massive than $10^{10}\msun$~\citep{kru12,kuh12}.  If true, then
the $z=6$ UV LF is expected to turn over at an absolute 
magnitude of $\approx-15$~\citep{jaa13}.  

Given the importance of the high-redshift UV LF as a probe of activity
in low-mass halos, multiple techniques for interpreting it have been 
developed.  One method involves assuming that local group dwarf galaxies 
are representative of the 
high-redshift population and using spatially-resolved star formation
histories to infer the abundance and activity in their progenitor 
population.~\citet{wei14b} used this approach to argue that the intrinsic 
$z>5$ LF grows without a turnover to at least $\MUV=-5$.  However, a 
subsequent analysis found that it may flatten for 
$\MUV>-13$ (\citealt{boy15}; see also~\citealt{boy14}),
in better agreement with predictions from ab-initio 
simulations~\citep[for example,][]{wis14}.

% Add references to Bouwens' work below
More directly, a number of groups have used imaging from the 
Hubble Frontier Fields to trace the UV LF out to unprecedented 
depths~\citep{ate15,ish15,ish16,lap16,liv16,bou16}.  
The deepest measurements are reported by~\citet{liv16}, who use a 
wavelet decomposition approach to remove foreground light from galaxies
associated with the lensing clusters.  This allows them to identify many 
more faint systems than had 
previously been detected.  They find that the UV LF is inconsistent with 
a turnover at $M<-12.5$ at $z=6$, with weaker constraints at higher 
redshifts.

What do the~\citet{liv16} measurements imply for galaxy growth in 
low-mass halos? Qualitatively, the finding that the slopes of the dark matter
halo mass function at low masses and the UV LF at faint luminosities
are similar ($\approx -2$ in both cases) implies that galaxy luminosity
varies nearly linearly with the host halo's mass down to the faintest
detected systems.  Moreover, evidence for a minimum halo mass below 
which gas accretion and star formation are inefficient has not yet been 
detected.  Instead, the~\citet{liv16} data place an upper limit on $\Mmin$.
This limit constrains the 
efficiency of physical effects that limit star formation in low-mass halos.  
Additionally, it informs simplified reionization models that ``paint" 
luminosities directly onto dark matter halos: they are disfavored if they 
invoke significant suppression where it is not 
observed.

A simple way to derive this limit is to impose a minimum halo mass cutoff 
on a galaxy formation model and ask how large that cutoff may be made before 
it introduces conflict with the observed LF.  While similar comparisons have 
been undertaken before~\citep{mun11,cas16}, the 
arrival of significantly deeper measurements motivates us to revisit the 
problem.  Additionally, galaxy formation models are now able to address 
a broader range of observables simultaneously than before.  
In particular,~\citet{fin16} discussed a numerical hydrodynamic + continuum 
radiation transport model that reproduces both the observed 
UV LF and the abundance of low-ionization metal absorbers~\citep{fin16}.
Moreover, it predicts an integrated optical depth to Thomson scattering of 
0.059, in excellent agreement with the most recent constraints from the 
cosmic microwave background~\citep{pla16a,pla16b}.  In short, this 
model yields favorable agreement with observations of galaxies, absorbers, 
and a spatially-inhomogeneous reionization history simultaneously, 
opening up the possibility of understanding how these observables 
relate to one another.  It is therefore a particularly well-tested 
framework for interpreting the observed LF.  

In this work, we use this simulation to interpret the constraints reported 
by~\citet{liv16} as an upper limit on $\Mmin$ at $z>6$--8, and on the 
lowest stellar mass of the galaxies that have been 
observed at this epoch.  In Section~\ref{sec:sims}, we review 
our simulations.  In Section~\ref{sec:res}, we present 
our results.  In Section~\ref{sec:disc}, we discuss their implications
and compare to previous work. In Section~\ref{sec:sum}, we summarize.

\section{Simulation and Analysis}\label{sec:sims}
We ran our cosmological radiation hydrodynamic simulation using a custom 
version of {\sc Gadget-3}~\citep{spr05}.  We previously presented this 
simulation in~\citet{fin16}; the reader is referred to that work and 
to Section 2 of~\citet{fin15} for details of the physical treatments.  
Briefly, we discretize the mass in a periodic, cubical volume of comoving 
length 
$7.5\hmpc$ into $2\times320^3$ dark matter and gas resolution elements 
and the radiation field into a regular spatial grid of $40^3$ voxels 
and 16 independent frequency bins.  Gas whose proper density exceeds
0.13 $\cm^{-3}$ forms stars via a subgrid multiphase treatment~\citep{spr03}.
Star-forming gas is selectively added to
galactic outflows following the ``ezw" model~\citep{dav13}. 
The ionizing emissivity is tied to the local star-forming gas particles'
metallicities and star formation rates using the 
{\sc Yggdrasil}~\citep{zac11} spectral synthesis code.  The assumed
ionizing escape fraction depends both on redshift and halo 
mass~\citep{fin15}.  The radiation field is evolved using a moment 
method, and we iterate at each timestep between the ionization and 
radiation solvers to obtain a consistent solution~\citep{fin11,fin12}.

We generate the initial conditions using an~\citet{eis99} power spectrum 
at $z=249$.  We compute the initial gas ionization and temperature 
using {\sc RECFAST}~\citep{won08} Our adopted cosmology is one in 
which $\Omega_M=0.3$, $\Omega_\Lambda=0.7$, 
$\Omega_b = 0.045$, $h=0.7$, $\sigma_8 = 0.8$, and the index of the 
primordial power spectrum  $n=0.96$.  

\begin{figure}
\centerline{
\setlength{\epsfxsize}{0.5\textwidth}
\centerline{\epsfbox{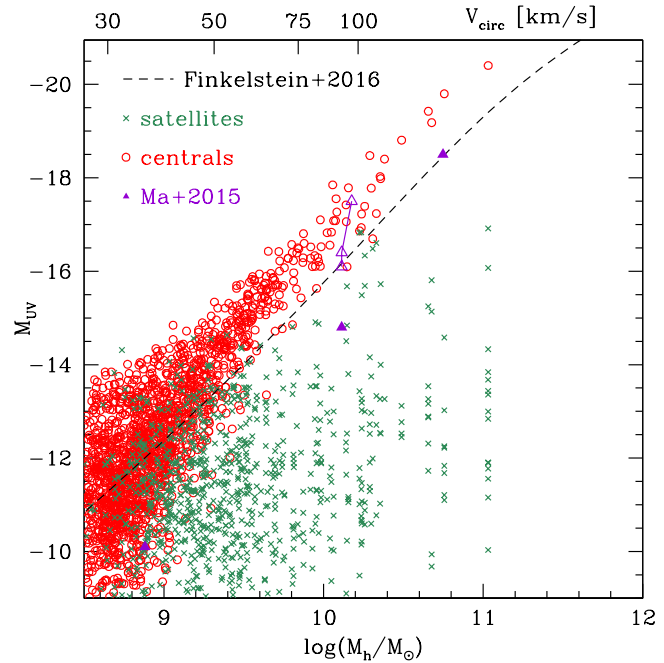}}
}
\caption{The simulated relationship between halo mass and UV luminosity
at $z=6$ both for centrals (red circles) and for satellites (green 
crosses).  The scatter is quite significant, particularly for 
$\MUV>-17$.
}
\label{fig:MUVMhalo}
\end{figure}

For each of our $z=6$, 7, and 8 snapshots, we identify simulated galaxies 
using {\sc skid}\footnote{http://www-hpcc.astro.washington.edu/tools/skid.html
} and compute their rest-frame 1500 \AA~ luminosities (in 
$\rm{ergs}\, \rm{sec}^{-1}\, \rm{Hz}^{-1}$) using
version 2.3 of the Flexible Stellar Population Synthesis 
library~\citep{con09}, interpolating to each star particle's age and 
metallicity. The summed luminosities for each model galaxy are then 
expressed as absolute AB magnitudes.  We neglect dust extinction because
galaxies at $z\geq6$ are observed to have relatively blue UV 
continua~\citep{bou14}.  In detail, our simulations do permit a small
amount of dust; we will return to this possibility in our discussion of 
Figure~\ref{fig:MUVMhalo} and consider it more closely in a 
future study.

We compute each galaxy's host halo mass by growing a sphere about its 
center of mass until it encloses an overdensity that matches the expected 
overdensity of collapsed systems.  A merging step re-assigns satellite
galaxies to parent halos; we find that roughly 40\% of galaxies are 
satellites.

Fundamentally, our study leverages the numerically predicted relationship 
between halo mass and UV luminosity, which we show in Figure~\ref{fig:MUVMhalo}.
The model predicts significant scatter in this relationship: the host halo
mass can vary by an order of magnitude even among centrals for $\MUV\leq-14$.
Consequently, a given minimum luminosity corresponds to a lower minimum
halo mass in our model than in a model in which the luminosity-halo mass
relation is assumed to be scatter-free.  

For context, we also compare our 
predicted halo mass-luminosity relation with 
two complementary analyses.  
~\citet{fink16} used an abundance-matching analysis
to derive a relationship that has the same slope, but for which a typical
halo is roughly one magnitude fainter (compare the circles with the dashed 
curve in Figure~\ref{fig:MUVMhalo}).  
This offset is somewhat surprising given that both relationships reproduce 
the observed UV LF.  Assuming that the~\citet{fink16} fit is
driven by bright galaxies, the inconsistency may confirm the suggestion
that the simulation overproduces the bright end of the 
observed LF (Figure~\ref{fig:lfz6}), though perhaps not as significantly
as one might expect from Figure~\ref{fig:MUVMhalo}.  The discrepancy cannot
be resolved by adding dust to our models without rendering their UV continuum
slopes too red: The predicted slopes $\beta$ (where the flux 
$F_\lambda\propto \lambda^\beta$) for sources with $\MUV\approx$ -17 -- -19
are characteristically -2.5, whereas observations indicate typical values
of -2.3~\citep{bou14}.  Adding enough dust to redden the predicted slopes
into agreement with observations would dim them by 
$\leq 0.5$ mag~\citep{cal00}, eliminating no more than half of the gap 
between our predictions and~\citet{fink16}.  We have also verified 
that our simulation's halo mass function agrees with analytical
expectations to within 0.1 dex computed using 
{\sc hmfcalc}~\citep{mur13}, so it is not the case that the model matches
the observed LF by underproducing halos and overproducing stars within 
each halo.  In any case, the disagreement at the fainter end, which is
the subject of this work, is weaker owing to the large predicted scatter.

We also compare our predictions to those of~\citet{ma15}, who recently 
studied a small number of simulated galaxies with unprecedented numerical
resolution and physical realism.  Using a similar analysis to ours (including the decision
to neglect dust), they find the relationship indicated by the solid magenta
triangles.  Their model predicts that galaxies are systematically 1--2 
magnitudes fainter at a given halo mass than ours, although the 
lowest-mass halo lies within our model's predicted scatter.  Consistently
with their Figure 4, we find that they predict overall fainter luminosities
at given halo mass than is inferred from abundance-matching analyses. 
\citet{ma15} also re-simulated their $10^{10}\msun$ halo at eight times lower mass 
resolution using three different density thresholds for star formation, 
leading to generally brighter galaxies (open triangles) that are in much 
better agreement with our model and, ironically, empirical inferences 
(as represented here by~\citealt{fink16}).  It is interesting to note that 
decreasing the mass resolution by a factor of 8 has a larger effect 
on $\MUV$ than changing the threshold density for star formation by a 
factor of 100.  This highlights the numerical challenge of modeling 
even individual galaxies at high redshift in a resolution-convergent 
way.

\section{Analysis and Results}\label{sec:res}
\begin{figure}
\centerline{
\setlength{\epsfxsize}{0.5\textwidth}
\centerline{\epsfbox{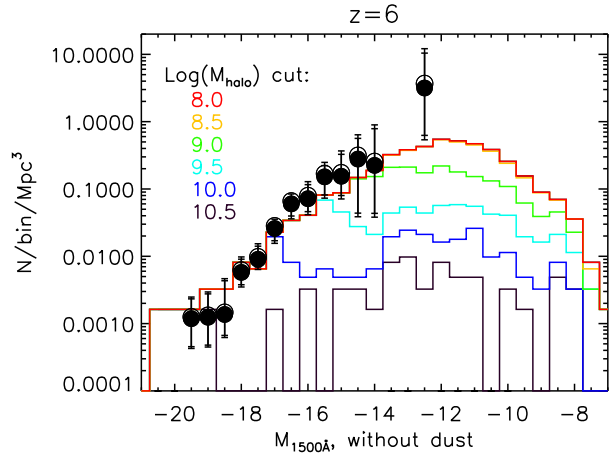}}
}
\caption{The simulated and observed luminosity functions at $z=6$.  As
the minimum dark matter halo mass is increased, successively brighter
galaxies are eliminated from the predicted LF.  Measurements are 
from~\citet{liv16} and are shown both without (solid) and with (open)
a correction for Eddington bias (see text).
}
\label{fig:lfz6}
\end{figure}

In Figure~\ref{fig:lfz6}, we illustrate our approach.  With no tuning of 
parameters, the predicted dust-free LF is in excellent agreement with 
observations at $z=6$ (red curve).  We next impose a turnover in the LF 
by removing galaxies hosted by dark matter halos below a variable mass 
threshold $M_c$.\footnote{For clarity, note that we use $M_c$ to denote the 
model parameter and $\Mmin$ as the physically relevant minimum halo mass
that we are trying to infer; that is, the value of $M_c$ that is preferred 
by the data is a constraint on $\Mmin$.}
Increasing $M_c$  moves the turnover to higher 
luminosity, weakening the agreement with 
observations.\footnote{Imposing $M_c=10^8\msun$ (red curve in Figure~\ref{fig:lfz6}) 
does not remove any star-forming halos in our simulation at $z\leq8$ 
owing to photoionization heating, so it is equivalent to $M_c=0$.}
In the case of satellite galaxies, we remove them only if their 
parent halo mass exceeds the threshold.  Qualitatively, this will lead 
to a \emph{higher} (and hence more conservative) limit on $\Mmin$ than 
if we only considered the subhalo mass.

For a given cutoff halo mass $M_c$, we compute a $\chi^2(M_c)$ statistic 
that quantifies the agreement between the predicted and observed LFs:
\begin{equation}\label{eqn:chisq}
\chi^2(M_c) \equiv \sum_{i=1}^N\frac{(\phi_{M,i}(M_c) - \phi_{D,i})^2}{\sigma_{D,i}^2}
\end{equation}
Here, $i$ runs over the $N$ observed magnitude bins; $\phi_{M,i}(M_c)$ and 
$\phi_{D,i}$ are the modelled and observed LFs in the $i$th magnitude bin, 
respectively, while $\sigma_{D,i}$ is
the reported uncertainty in the $i$th magnitude bin.  As the reported errors 
are slightly asymmetric, we use the (upper, lower) error if the model is 
(above, below) the data.  With no tuning of parameters, $\chi^2(M_c=0)=9.91$ 
for 13 data points, reinforcing the visual impression that the baseline 
model is in good agreement with the data.

Having computed $\chi^2(M_c)$, we follow~\citet{liv16} by using the Bayesian 
Information Criterion (BIC) to ask how strongly a given minimum halo mass is ruled 
out.  First, we compute the BIC for the base model, which is the simulation 
itself with no adjustable parameters; this is just $\chi^2(M_c = 0)$.  Next,
we vary the cutoff halo mass $M_c$ and compute the BIC
\begin{equation}\label{eqn:BIC}
\rm{BIC}(M_c) = \chi^2(M_c) + k\ln(N)
\end{equation}
where $k=1$ is the number of parameters in the adjustable model and $N$ is the
number of magnitude bins for which~\citet{liv16} report a LF measurement.
We then compute $\Delta\rm{BIC}(M_c) \equiv \rm{BIC}(M_c) - \rm{BIC}(0)$. 
\citet{kas95} find that $\Delta\rm{BIC}>(2,6,10)$ indicates (positive, strong,
very strong) evidence against the more adjustable model.

\begin{figure}
\centerline{
\setlength{\epsfxsize}{0.5\textwidth}
\centerline{\epsfbox{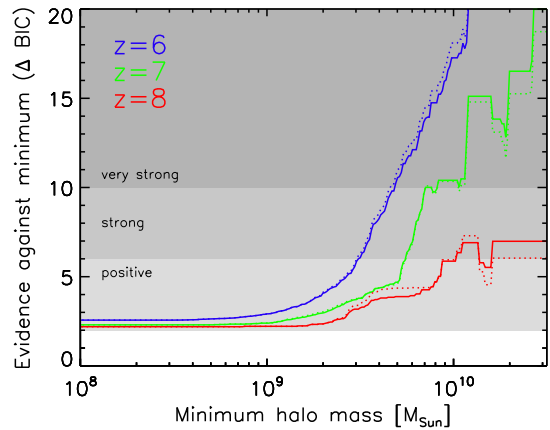}}
}
\caption{$\Delta\rm{BIC}(M_c)$ versus $M_c$; (blue, green, red) indicate the constraints at $z=$(6,7,8).  Solid/dashed curves indicate inferences without/with a correction to the data for Eddington Bias.  The data, when compared with our model, provide strong evidence against models in which star formation is suppressed in halos less massive than $10^{9.5}\msun$ at $z=6$, with weaker constraints resulting at higher redshifts.
}
\label{fig:BIChalo}
\end{figure}

At $z=6$, we find that $\Delta\rm{BIC}>2$ for all adjustable models simply
because $N>10$.  However, it exceeds (6,10) for $\log(M_c) > (9.5,9.7)$,
indicating strong evidence against significant suppression of star formation
in halos less massive than $10^{9.5}\msun$ (Figure~\ref{fig:BIChalo}). 
Repeating this analysis at $z=7$, we find that $\Delta\rm{BIC}$ is above
(6,10) for $\log(M_c) = (9.75,9.95)$. Models in which 
$\Mmin\geq10^{10}\msun$ are 
therefore ruled out for $z=6$--7.  At $z=8$, the shallower nature of 
the data leads to a weaker constraint: $\Delta\rm{BIC}>6$ for 
$\log(M_c) > 10.25$, and it never exceeds 10.
We have separately evaluated the impact of satellites by ignoring them 
entirely, finding that the inferred threshold halo mass decreases by 
less than a factor of two.  For simplicity, however, we report results 
in the case where satellite systems are included.

A related question regards the minimum stellar mass that is currently probed
by Frontier Fields data: Are they sensitive to analogues to the progenitors of
Local Group dwarf galaxies? The answer depends both on the minimum
stellar mass that the observations probe, and on the amount of subsequent 
growth that can be expected from $z=6\rightarrow0$.  We address this question 
using the same approach as before: by removing galaxies whose total formed
stellar mass exceeds a threshold and computing the $\Delta$BIC that
relates the predicted luminosity functions with and without the threshold.
Throughout this analysis, we adopt the total stellar mass formed rather than
the mass that remains in stars in order to facilitate comparison 
with~\citet{wei14a}.

\begin{figure}
\centerline{
\setlength{\epsfxsize}{0.5\textwidth}
\centerline{\epsfbox{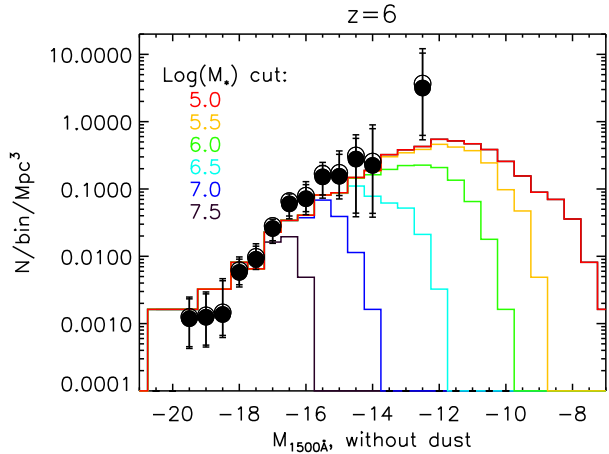}}
}
\caption{The simulated and observed luminosity functions at $z=6$.  The red
curve shows the LF if no simulated galaxies are omitted; it is the same as the 
red curve in Figure~\ref{fig:lfz6}.  Analogously to Figure~\ref{fig:lfz6}, 
increasing the minimum stellar mass removes successively brighter galaxies, 
weakening the agreement with observations.
}
\label{fig:lfz6star}
\end{figure}

\begin{figure}
\centerline{
\setlength{\epsfxsize}{0.5\textwidth}
\centerline{\epsfbox{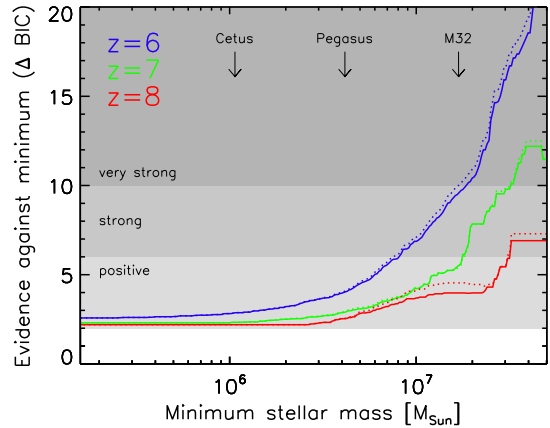}}
}
\caption{$\Delta\rm{BIC}(M_*)$ versus $M_*$; (blue, green, red) indicate the constraints at $z=$(6,7,8); the size of the Eddington Bias correction is indicated as in Figure~\ref{fig:BIChalo}.  The data, when compared with our model, provide strong evidence that observations probe at least down to stellar masses of $6\times10^{6}\msun$ at $z=6$, with higher minima at higher redshifts.  The arrows label the total stellar masses formed in three local group dwarf galaxies by $z=5$~\citep{wei14a}.  Current data likely probe the progenitors of massive dwarfs such as Pegasus and M32, but not smaller ones such as Cetus.
}
\label{fig:BICstar}
\end{figure}

As we show in Figure~\ref{fig:lfz6star}, the level of agreement with observations
at $z=6$ begins to weaken once the minimum stellar mass is increased above 
$10^6\msun$.  Quantifying this impression via the $\Delta\mathrm{BIC}$ analysis, 
we find that models in which observations do not probe down to stellar masses of 
$8.1\times10^6\msun$ at $z=6$ are strongly disfavored (Figure~\ref{fig:BICstar}).  
The minimum stellar mass is somewhat larger at higher redshifts, but even at
$z=8$ we find that current observations probe down to stellar masses of 
$3.2\times10^7\msun$.

To place this scale in context, we have used~\citet{wei14a} to compute the
total stellar mass formed in various local group dwarfs by $z=4.8$.  These
range from $\sim10^3$--$10^7\msun$, with the implication that some of them
were already massive enough at $z=5$ that the Frontier Fields may be
identifying analogues to their progenitors at $z\geq6$.  To emphasize this
overlap, we indicate the total stellar mass formed in three representative
dwarfs by $z=4.8$ in Figure~\ref{fig:BICstar}.  The progenitors of M32 
and Pegasus would very likely have been observable, while that of Cetus 
would not.  Deeper
\emph{HST} data will be required to constrain the amount of star 
formation that occurred in local dwarfs prior to $z=4.8$, and will test 
the association between local group and Frontier Fields dwarfs in more 
detail.

The~\citet{liv16} LF measurements may be subject to Eddington bias; that is, 
the tendency to overestimate the proportion of rarer, brighter objects owing to 
contamination by more common, fainter objects that scatter into brighter bins.
~\citet{liv16} account for this consistently when they fit Schechter functions 
to their measurements, but the measurements themselves are uncorrected.  In 
order to account for this, we compute the ratio of the Schechter functions 
derived both with and without the Eddington bias 
correction\footnote{The uncorrected fits are not published} and take the 
ratio of these as a correction factor.  The correction varies with magnitude 
but is in the range of 0.8--0.9 for $z=$6--7 and 0.5--0.9 for $z=8$.  
We then derive the $\Delta$BIC curve both with (dashed) and without 
(solid) the correction.  Correcting 
only the measurements in this way (while leaving the reported uncertainties 
unchanged) lowers the LF amplitude, increasing the limit on $\Mmin$ by
0.05--0.1 dex.  Intuitively, if there are fewer galaxies observed, then we 
may remove fewer halos from the model before it conflicts with data.  However,
multiplying the uncertainties by the same correction factor makes 
$\chi^2(M_c)$ more sensitive to $M_c$, moving the $\Delta$BIC curve to lower
halo masses by a similar amount.  The end result is a nearly-complete 
cancellation between the two corrections (compare the dashed and solid curves).
We conclude that Eddington bias is not a major uncertainty in the minimum
halo and stellar mass of the systems that~\citet{liv16} identify.

In summary, the~\citet{liv16} data provide strong evidence against 
significant suppression of star formation in halos less massive than 
$10^{10.25}\msun$ throughout the range $z=$6--8, with the strongest 
constraint coming from $z=6$, where the data require star 
formation in halos at least as massive as $10^{9.5}\msun$.  At the
same time, our model suggests that the faintest directly-detected 
galaxies are comparable in stellar mass to the progenitors of Local 
Group dwarfs.  This supports
the exciting possibility of using reionization-epoch observations 
to study the early stages of their growth, just as it supports 
using the local group to draw inferences regarding the galaxies that
drove reionization~\citep[for example,][]{boy14}.
%FIXME: Continue editing here...

\section{Discussion}\label{sec:disc}

The purpose of this study is, in essence, to invoke the relation shown in 
Figure~\ref{fig:MUVMhalo} (and its analogues at higher redshifts) 
in order to constrain $\Mmin$ from the observed
UV LF.  Our analysis shows that current observations already exert
pressure on some previous treatments for star formation.  For example, 
we find that star formation must be relatively unsuppressed in halos 
down to $3.1\times10^9\msun$ at $z=6$, whereas the fiducial 
metallicity-based H$_2$ model of~\citet{kru12} predicts $\approx80\%$ 
suppression at this mass scale (their Figure 8).  
Our results are also in conflict with the accretion-floor model of~\citet{bou10}: if
halos below $10^{11}\msun$ do not form stars, then galaxies must populate
halos in a very different way than arises in hydrodynamic 
simulations (that is, Figure~\ref{fig:MUVMhalo}) in order to match 
the~\citet{liv16} measurements.  Of course, the~\citet{bou10} model 
was forwarded as an interpretation of observations at $z\leq6$, hence 
a more convincing test would be to repeat our analysis using the 
observed UV LF at lower redshifts.  For the present, we therefore
limit ourselves to the conclusion that an accretion floor at 
$10^{11}\msun$ does not apply at $z>6$.  This agrees with~\citet{beh13}, 
whose analysis of the stellar mass - halo mass relation likewise 
indicates robust star formation in $10^{10}\msun$ halos out to $z=8$.

An obvious improvement over our study would be to 
\emph{test} the prediction presented in Figure~\ref{fig:MUVMhalo}.  
In order to motivate the need for such a test, we compare our analysis
to two previous efforts.~\citet{mun11} constructed a semi-analytical 
model and used it to analyze the shallower observations that were 
available at the time ($\MUV \leq -18$).  In their entirely 
complementary model,
halos are assumed to condense their gas into a star-forming disk
whenever there is a merger, after which the condensed gas forms 
into stars.  They fitted simultaneously for the luminosity amplitude 
$L_{10}$ (that is, the luminosity $\log(L_{1500}/\erg\rm{\ s}^{-1} \Hz^{-1})$ 
within $10^{10}\msun$ halos), and for $\Mmin$ ($m_\mathrm{supp}$ in
their notation), and found that $\Mmin\leq10^{9.5}\msun$ at $z=6$ (we will 
compare with their results only at $z=6$; similar results occur at 
$z=7$ and $z=8$).  In other words, for an observed luminosity function that 
probes $100\times$ shallower than~\citet{liv16}, they derived roughly the 
same $\Mmin$ as we do.  It is reasonable to assume that, were they to 
confront their model with the most recent constraints, the inferred 
$\Mmin$ would be much lower than ours.  This discrepancy 
reveals a very significant theoretical uncertainty.

The fact that both models match the observed LF's normalization 
despite this remarkable discrepancy owes to a cancellation between two 
effects.  First, their derived normalization $L_{10}=27.2$ is nearly 
$10\times$ larger than our model, which predicts $L_{10}=26.3$ 
(Figure~\ref{fig:MUVMhalo}), corresponding
to a much lower star formation efficiency.  Second, their model
predicts highly bursty star formation histories: For their fiducial
model, the ``active fraction" of $10^{10}\msun$ halos is $<50\%$.
Additionally, their model assumes that star formation in satellite 
halos stops whenever there is a major merger.  This contrasts with 
our model, in which star formation is generally smooth~\citep{fin11}
and satellites constitute a significant fraction of the observable
population (Figure~\ref{fig:MUVMhalo}).  Together, these differences 
increase the predicted characteristic luminosity at a given halo 
mass with respect to our model.  

More recently,~\citet{cas16} (see also~\citealt{yue14,yue16}) confronted 
a different semi-analytic
model with Frontier Fields measurements that probed down to $\MUV=-15$
and inferred that the threshold circular velocity must be below
$50\kms$, corresponding to a halo mass threshold of 
$\log(M_c/\msun)=9.3$.  Given that their input LF is 2--3 magnitudes 
shallower than ours, the result that the inferred halo mass 
threshold is similar once again implies a discrepancy in the 
underlying physics.  It is likely that, as before, tradeoffs 
between the unknown burstiness and star formation efficiency of
high-redshift galaxies are to blame. 

The need for observations that can distinguish between these models 
is clear.  For example, \emph{JWST} will enable measurements of
the relationship between complementary observables such as UV
luminosity, continuum slope, stellar mass, and H$\alpha$ luminosity,
which will probe the underlying level of burstiness.  In the nearer
term, comparison with clustering measurements may already be able
to distinguish between the models; however, this is beyond the
scope of the current study.  For the present, it is most noteworthy 
that, in all three analyses, observations forbid star formation to 
be suppressed in halos above $\log(M_c/\msun)=9.5$.

The uncertainties inherent in using galaxy formation models to interpret 
the UV LF may be mitigated by invoking observations of long gamma-ray 
bursts (LGRBs), which are also believed to trace star 
formation.~~\citet{wei16} recently used a simple model that ties galaxy 
growth and LGRBs to the evolving dark matter halo mass function to derive 
$\Mmin$ from \emph{Swift} observations.  At $1\sigma$ confidence, they 
constrain it to be less than $10^{10.5}\msun$ at $z<4$ and in the range of 
$10^{7.7}$--$10^{11.6}\msun$ at $4 < z < 5$.  These results are 
consistent with ours, although they apply to lower redshifts.

Finally, we note that there is room for progress in reducing both 
observational and theoretical uncertainties.  On the observational side, 
the luminosity of the faintest \emph{Hubble Frontier Fields} sources 
remain uncertain owing to the challenge of measuring faint sources in 
lensed fields.  Additionally, the unknown intrinsic sizes of the lensed 
sources introduces uncertainty into the observational incompleteness 
corrections that are needed to compute volume densities.  The effect
is particularly dramatic for the faintest luminosities~\citep{bou16}.
On the theoretical side, resolution limitations may affect our analysis: 
with our cosmology and dynamic range, the 
minimum stellar mass to which~\citet{liv16} probe ($6\times10^6\msun$; 
Figure~\ref{fig:BICstar}) corresponds to $\approx52$ star particles, whereas 
we have previously argued that 64 are required for converged predictions 
of global galaxy properties such as mass and luminosity~\citep{fin06}.  
The host halos, by contrast, are quite well-resolved: the minimum halo mass of 
$3.1\times10^9\msun$ corresponds to 7100 dark matter particles, which is 
more than sufficient to resolve both the halo's mass and internal 
structure~\citep{tre10} as well as its gas accretion history~\citep{nao09}.  
Hence while we do not believe that resolution limitations are severe, at 
present they limit our ability to comment in more detail on the nature 
of the faintest currently-observed galaxies.

\section{Summary}\label{sec:sum}
We have combined predictions from a cosmological radiation hydrodynamic 
simulation with observations of the UV LF at $z=$6--8 in order to constrain
the minimum mass dark matter halo in which star formation is unsuppressed.
We find that recent observations require vigorous star formation in
halos at least as massive as (3.1, 5.6, 10.5)$\times10^9\msun$ at 
$z=(6,7,8)$, ruling out models in which significant suppression is expected
in halos as massive as $10^{10}\msun$.  Likewise, we find that these
observations probe objects with total formed stellar masses (at the observed epoch)
in the range 8--32$\times10^6\msun$.  This overlaps with the range of 
inferred stellar masses of local group dwarfs at $z=4.8$, indicating
that the Frontier Fields may well contain fairly direct insights into
the early growth histories of local dwarfs.  Lingering degeneracies 
between unknowns such as the duty cycle of star formation and the 
normalization of the $\MUV$-$M_h$ relation indicate that future work 
involving galaxy colors and clustering measurements are required in order
to constrain models further.

\section*{Acknowledgements}
We thank our colleagues at NMSU for helpful conversations and input and 
R.\ Bouwens for comments on an early draft.  We also thank the referee for
comments that improved the draft.  We are indebted to V.\ Springel 
for making {\sc Gadget-3} available to our group, and to P.\ Hopkins for 
kindly sharing his SPH module.  Our simulation was run on {\sc Gardar}, 
a joint Nordic supercomputing facility in Iceland.
KF thanks the Danish National Research Foundation for funding the Dark Cosmology Centre.  
EZ acknowledges research funding from the Swedish Research Council, the Wenner-Gren 
Foundations and the Swedish National Space Board.

\end{document}